\documentclass[floats, prd, showpacs, nofootinbib,twocolumn]{revtex4-1}

\usepackage{color,graphicx}
\usepackage{amsfonts}
\usepackage{amssymb}
\usepackage{comment}
\begin{document}

\title{Can accretion properties distinguish between a naked singularity, wormhole and black hole?}
\author{R.Kh. Karimov${}^{1}$}\email{karimov\_ramis\_92@mail.ru}
\author{R.N. Izmailov ${}^{1}$}\email{izmailov.ramil@gmail.com}
\author{A.A. Potapov ${}^{2}$}\email{a.a.potapov@strbsu.ru}
\author{K.K. Nandi${}^{1,2,3}$}\email{kamalnandi1952@rediffmail.com}

\affiliation{
${}^{1}$Zel'dovich International Center for Astrophysics, Bashkir State Pedagogical University, 3A, October Revolution Street, Ufa 450008, RB, Russia \\
${}^{2}$Department of Physics \& Astronomy, Bashkir State University, 47A, Lenin Street, Sterlitamak 453103, RB, Russia \\
${}^{3}$High Energy Cosmic Ray Research Center, University of North Bengal, Darjeeling 734 013, WB, India \\
}

\begin{abstract}
We first advance a mathematical novelty that the three geometrically and topologically distinct objects mentioned in the title can be exactly obtained from the Jordan frame vacuum Brans I solution by a combination of coordinate transformations, trigonometric identities and complex Wick rotation. Next, we study their respective accretion properties using the Page-Thorne model which studies accretion properties exclusively for $r\geq r_{\text{ms}}$ (the minimally stable radius of particle orbits), while the radii of singularity/ throat/ horizon $r<r_{\text{ms}}$. Also, its Page-Thorne efficiency $\epsilon$ is found to increase with decreasing $r_{\text{ms}}$ and also yields $\epsilon=0.0572$ for Schwarzschild black hole (SBH). But in the singular limit $r\rightarrow r_{s}$ (radius of singularity), we have $\epsilon\rightarrow 1$ giving rise to $100 \%$ efficiency in agreement with the efficiency of the naked singularity constructed in [10]. We show that the differential accretion luminosity $\frac{d\mathcal{L}_{\infty}}{d\ln{r}}$ of Buchdahl naked singularity (BNS) is always substantially larger than that of SBH, while Eddington luminosity at infinity $L_{\text{Edd}}^{\infty}$ for BNS could be arbitrarily large at $r\rightarrow r_{s}$ due to the scalar field $\phi$ that is defined in $(r_{s}, \infty)$. It is concluded that BNS accretion profiles can still be higher than those of regular objects in the universe.
\end{abstract}

\maketitle

\section{Introduction}
\label{intro}

Astrophysical observations suggest that almost all black holes (BH) are surrounded by gas clouds with accretion disks of various sizes, extending from about ten to hundred parsecs \cite{Urry:1995}. However, BHs need not be the only accreting objects in the sky. There could be other categories of hypothetical objects such as naked singularity (NS) and wormholes (WH) that are not ruled out either by theory or by experiment to date. On the contrary, there has been a recent surge of interest in the study of their observational signatures, which include, but not limited to, the phenomena of accretion
\cite{Broderick:2007,Harko:2008,Harko:2009a,Harko:2009b,Kovacs:2010,Chowdhury:2012,Bambi:2013a,Bambi:2013b,Joshi:2014,Blaschke:2016,Karimov:2018,Karimov:2019,Shaikh:2019a,Gyulchev:2019}, gravitational lensing \cite{Virbhadra:2002,Nandi:2006,Gyulchev:2008,Dey:2008,Bhattacharya:2010,Nakajima:2012,Tsukamoto:2016,Tsukamoto:2017,Shaikh:2017,Lukmanova:2018,Jusufi:2018,Nandi:2018,Izmailov:2019a,Izmailov:2019b,Shaikh:2019b,Shaikh:2019c}, shadow cast on the background of the thin accretion flow \cite{Lacroix:2013,Johannsen:2016a,Johannsen:2016b,Nedkova:2013,Ohgami:2015,Abdujabbarov:2016,Gyulchev:2018,Shaikh:2018,Shaikh:2019d,Amir:2019} and gravitational waves \cite{Konoplya:2016,Cardoso:2016a,Cardoso:2016b,Cardoso:2017,Nandi:2017,Bueno:2018,Volkel:2018}. These studies
offer a wealth of information about these objects but the individual spacetimes describing them are chosen either absolutely freely or constructed artificially and as a rule they cannot be analytically derived from one another by mathematical algorithms. We shall show here that there could be a novel exception to this rule -- some famous static spherically symmetric asymptotically flat NS and WH solutions can indeed be derived from one another by a combination of coordinate transformations, trigonometric identities and complex Wick rotation. The natural question then is to ask if these solutions, despite being mathematically non-trivially connected, can nevertheless be observationally distinguishable. To answer this question, we shall investigate the diagnostic of their accretion properties. Among other things, the two major takeaways from the present investigation are: (1) The possibility of conversion of NS into an \textit{everywhere} regular asymptotically flat WH solution and vice versa, while both lead to Schwarzschild BH (SBH) as a special cases. (2) While BH is undoubtedly the most favored candidate for the accreting astrophysical object, the profiles of other competing objects such as those of NS and regular traversable WH cannot be observationally ruled out as yet.

To elaborate what we mean, consider vacuum Jordan frame (JF) Brans-Dicke theory in the JF with the scalar field $\varphi$ playing the role of spin$-0$ gravity. Originally Brans \cite{Brans:1962a} listed four separate classes of matter-free solutions (I-IV) in the Machian JF of which only the first two are asymptotically flat. JF Brans I solution represents a NS due to the appearance of a curvature singularity at a finite non-zero coordinate radius. The solution can also be interpreted as a singular non-traversable WH connecting two asymptotically flat spacetimes across the singularity for a narrow range of the Brans-Dicke coupling constant $\omega$ \cite{Nandi:1998}. JF Brans I solution never represents BHs for finite values of $\omega $, as has been recently argued by Faraoni \textit{et al.} \cite{Faraoni:2019}. However, JF Brans I solution yields JF Brans II solution under some redefinitions and complex Wick rotations (see Sec.2), the latter truly being a twice asymptotically flat everywhere regular traversible WH \cite{Bhattacharya:2009}. Although geometrically and topologically quite different, JF Brans I and II are then seen to be \textit{not} strictly independent solutions as one can be derived from the other - an intriguing fact that seemingly went unnoticed by Brans himself and later researchers. Both Brans I and II are characterized by the same number of constants and reduce to SBH in the general relativity (GR) limit $\omega \rightarrow\infty $\footnote{%
The limit is not as simple as it looks, there are conceptual twists and turns, see e.g., \cite{Faraoni:1998,Faraoni:1999,Bhadra:2001}.}. The collapse of vacuum JF Brans I solution was shown by numerical simulation to lead to SBH with the scalar field radiated away \cite{Scheel:1995a,Scheel:1995b}.

In the conformally rescaled Einstein frame (EF), the equations are just those of GR without $\omega$ with the transformed scalar field $\phi$ playing the role of a material source term in the GR equations. Clearly, $\varphi$ in JF and $\phi$ in EF play completely diffrent roles. Both JF Class I and II solutions can be expressed in terms of a single arbitrary constant $\gamma$ in the EF (see Sec.2) but they still carry forward their JF characteristics into the EF. As long as $\gamma$ is kept real and $<1$, the JF Brans I NS converts in the EF to Buchdahl NS (BNS) \cite{Buchdahl:1959}, rediscovered a decade later now famously known as the Janis-Newman-Winnicour (JNW) solution \cite{Janis:1968}. Under some other transformations including Wick rotation, BNS converts in the EF to a regular, twice asymptotically flat traversible WH, which we call\ the Ellis-Bronnikov wormhole (EBWH) characterized by a real $\gamma$. \textit{We shall assume that accretion is taking place only on the attractive positive mass side ( }$0<r^{\prime }<\infty $\textit{) and so by the term EBWH in this paper, we shall mean only this form of the metric, i.e., Eqs.(36-39).}This coordinates can be extended to cover the full-patch ($-\infty <\ell <+\infty$), which then assumes the more familiar form of EBWH \cite{Ellis:1973,Bronnikov:1973} but we shall not use this form for our purposes. Like their JF predecessors, the two EF re-incarnations, BNS and EBWH, are also not independent solutions -- one can be obtained from the other, to be shown in Sec.2. However their passages to SBH differ: The BNS yields the SBH at the value $\gamma =1$, while EBWH yields the SBH at the value $\gamma =i$ \cite{Nandi:2017}. In either case, the scalar field $\phi$ vanishes. This means that, for collapse from EBWH to SBH, the constant $\gamma$ has to suddenly jump from a real value to an imaginary value $-i$ at the end of collapse auguring a topology change from a centerless and horizonless WH to a BH with a singular center ($r=0$) covered by a horizon.

A caveat should be mentioned before proceeding further. Our analysis presupposes the existence or stability of the considered solutions though these issues are not the object of this paper. The existence of NS is forbidden by the Penrose cosmic censorship conjecture (unproven) and the EBWH has been shown to be unstable under general perturbations collapsing into a SBH (see, e.g., \cite{Gonzalez:2009a,Gonzalez:2009b,Shinkai:2002,Bronnikov:2011}). On the other hand, under a restricted class of scalar field perturbations that vanish at the throat, EBWH has been shown to be stable \cite{Armendariz:2002}. Another recent new idea is that, due to curved spacetime, observation of instability or otherwise of EBWH could be \textit{perception-dependent} depending on the location of the observer - while one observer perceives instability, another observer might perceive stability and conversely (see, for details, \cite{Nandi:2016}). In any case, we do not contend the conclusion of instability but, instead of summarily ruling out EBWH, we advocate that verifiable diagnostics such as the accretion profiles (dealt with here), emission of gravitational waves, strong field lensing etc should be treated as observable verifications of EBWH, if there exists any in the universe \cite{Konoplya:2016,Nandi:2017}.

The purpose of this paper is to explicitly show how the above famous solutions, Brans I, BNS, EBWH and SBH, are connected with one another and to study the kinematic and emissivity properties of accretion around the last three objects. We then wish to investigate how, despite being connected across the real/imaginary divide of $\gamma$, the profiles differ from each other. We shall employ Page-Thorne model \cite{Page:1974} for accretion and assume for numerical illustration a toy model of a central object with mass $15M_{\odot}$ and accretion rate $\dot{M}=10^{18}$ gm.sec$^{-1}$, which could be identified with BNS, EBWH or SBH respectively.

In Sec.2, we explain the diagram showing the conversions among different solutions and Sec.3 shows the massless corollaries. Sec.4 describes the thin accretion disk with its kinematic and emissivity formulas belonging to the Page-Thorne model. Sec.5 analyses the accretion profiles of three objects, the BNS, EBWH and SBH. Sec.6 calculates the Eddington luminosity caused by the scalar field in the BNS and EBWH spacetimes. Sec.7 concludes the paper. We take $c=1,8\pi G=1$ unless specifically restored.

\section{Conversions among different solutions}
\label{sec:2}
The diagram below transparently tells the whole story of inter-connectibility of metrics. The arrows are pointed both ways to indicate that the transformation for going from one to the other metric can be reversed to return to the starting metric.

\begin{figure}[!ht]
  \centerline{\includegraphics[scale=1.8]{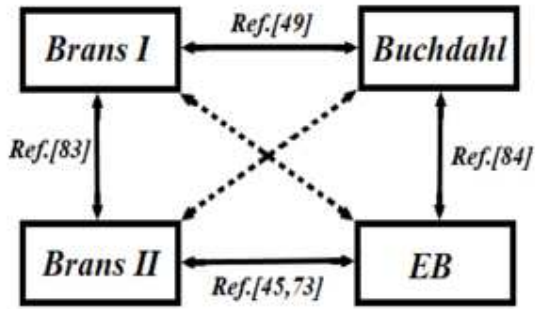}}
  \caption{The inter-convertibility of different metrics.}
  \label{Veff}
\end{figure}

\subsection{JF Brans I $\longleftrightarrow$ EF BNS (or JNW)}
The field equations obtained by varying the vacuum (matter free) JF Brans-Dicke action
\begin{equation}
S=\frac{1}{16\pi}\int d^{4}x(-g)^{\frac{1}{2}}\left[\varphi\mathbf{R+}\varphi^{-1}\omega g^{\rho\sigma}\varphi_{,\rho}\varphi_{,\sigma}\right]
\end{equation}%
which are
\begin{eqnarray}
\mathbf{R}_{\rho\sigma} - \frac{1}{2}g_{\rho\sigma}\mathbf{R} &=& -\frac{\omega}{\varphi^{2}}\left[\varphi_{,\rho}\varphi_{,\sigma}-\frac{1}{2}%
g_{\rho\sigma}\varphi_{,\eta}\varphi^{,\eta}\right] \nonumber \\
&&-\frac{1}{\varphi}\left[\varphi_{,\rho;\sigma}-g_{\rho\sigma}\square^{2}\varphi\right], \\
\square^{2}\varphi &=& 0,
\end{eqnarray}%
where $\square ^{2}\equiv (\varphi ^{;\chi })_{;\chi}$ and $\omega$ is a dimensionless coupling constant. The general solution of these field equations, in isotropic coordinates ($t,r,\theta ,\psi $), is
\begin{equation}
d\tau _{\text{{Brans I}}}^{2}\text{ }=-e^{2\alpha (r)}dt^{2}+e^{2\beta(r)}[dr^{2}+r^{2}(d\theta ^{2}+\sin ^{2}\theta d\psi ^{2})].
\end{equation}

The JF \textit{Brans I} solution \cite{Brans:1962a} is given by
\begin{eqnarray}
e^{\alpha (r)} &=& e^{\alpha _{0}}\left[ \frac{1-\frac{m}{2r}}{1+\frac{m}{2r}}\right] ^{\frac{1}{\lambda}}, \\
e^{\beta (r)} &=& e^{\beta _{0}}\left[ 1+\frac{m}{2r}\right] ^{2}\left[ \frac{1-\frac{m}{2r}}{1+\frac{m}{2r}}\right] ^{\frac{\lambda -C-1}{\lambda}}, \\
\varphi_{\text{Brans I}} &=& \varphi_{0}\left[ \frac{1-\frac{m}{2r}}{1+\frac{m}{2r}}\right]^{\frac{C}{\lambda}}, \\
\lambda^{2} &\equiv& (C+1)^{2}-C\left( 1-\frac{\omega C}{2}\right) >0,
\end{eqnarray}%
where $\lambda $, $m$, $\alpha _{0}$, $\beta _{0}$, $C$, and $\varphi _{0}$ are real constants, and the radial marker $r\in (m/2,\infty )$. The constants $\alpha _{0}$ and $\beta _{0}$ are determined by asymptotic flatness at $r=\infty $ as $\alpha _{0}=$ $\beta _{0}=0$. The negative $r-$side needs a bit of explanation. Note that the metric is invariant under inversion $\overline{r}\rightarrow \frac{1}{r}$ for even values of the exponents determined by $C(\omega )$, hence $\overline{r}=0$ is a second asymptotically flat region. Defining a new coordinate chart there by $\overline{r}\rightarrow \frac{1}{\widehat{r}}$, we find that $\widehat{r}\rightarrow -\infty $ represents the second asymptotically flat region with the radial markar $\widehat{r}\in (-\infty ,m/2$). Unfortunately, the two asymptotically flat spacetimes are disconnected by the NS at $r=\widehat{r}=\frac{m}{2}$, hence the solution cannot be accepted as a traversable WH. The surface $r=\widehat{r}=\frac{m}{2}$ is singular since curvature invariants diverge at that radius and is naked because the surface is not covered by a horizon. When $\lambda =1$ or $C=0$, we recover SBH in isotropic coordinates and the surface then becomes a regular horizon.

The Arnowitt-Deser-Misner (ADM) mass $M$ is given by%
\begin{equation}
M=\frac{1}{16\pi }\int_{S}\sum_{i,j=1}^{3}\left( \partial
_{j}g_{ij}-\partial _{i}g_{ii}\right) n^{i}dS,
\end{equation}%
where $S$ is a 2-surface enclosing the active gravitational region and $n^{i}$ is the unit outward normal. The Keplerian mass $M_{\text{Kepler}}$, which is also called the tensor mass \cite{Scheel:1995a,Scheel:1995b}, is obtained from the post-post-Newtonian (PPN) expansion of the metric. In the case of the JF\ Brans I solution (4), they yield, respectively
\begin{equation}
M_{\text{ADM}} = \frac{m(C+1)}{\lambda },\text{ }M_{\text{Kepler}}=\frac{m}{\lambda}.
\end{equation}

Under the conformal transformation%
\begin{equation}
\widetilde{g}_{\rho\sigma} = pg_{\rho\sigma},\text{ \ \ }p=\frac{1}{16\pi }%
\varphi ,
\end{equation}%
and a redefinition of the Brans-Dicke scalar%
\begin{equation}
d\phi =\left( \omega +\frac{3}{2}\right)^{1/2}\frac{d\varphi }{\varphi }.
\end{equation}%
Then the action (1) in the Einstein frame ($\widetilde{g}_{\rho\sigma},\phi $) becomes $\kappa \phi _{,\rho }\phi _{,\sigma }$%
\begin{equation}
S_{\text{EF}}=\int d^{4}x(-\widetilde{g})^{1/2}\left[ \widetilde{\mathbf{R}}+\widetilde{g}^{\rho \sigma }\phi _{,\rho }\phi _{,\sigma }\right] .
\end{equation}%
The field equations are%
\begin{eqnarray}
\widetilde{\mathbf{R}}_{\rho \sigma } &=&-\phi _{,\rho }\phi _{,\sigma}, \\
\square ^{2}\phi &=&0.
\end{eqnarray}

The right hand side represents ordinary matter if $\phi _{,\rho }\phi_{,\sigma }>0$ and ghost matter if $\phi _{,\rho }\phi _{,\sigma }<0$ (that violates Null Energy Condition). The solutions of Eqs.(14) and (15) can be obtained, using the transformations (11) and (12) on the Brans I metric (4), which we call here the Buchdahl solution \cite{Buchdahl:1959} in isotropic coordinates%
\footnote{%
There is a bit of history here. Originally, it was Fisher \cite{Fisher:1948}, who derived the solution for the canonical \textit{real} scalar field $\phi $ so that the stress tensor is positive, $\phi _{,\rho }\phi _{,\sigma }>0$. Since then it has been rediscovered under different parametrizations by other authors such as Bergman \& Leipnik \cite{Bergman:1957}, Buchdahl \cite{Buchdahl:1959} and Janis-Newman-Winnicour (JNW) \cite{Janis:1968}, to name a few. The equivalence of the last two are shown in \cite{Bhadra:2001b,Svitek:2016}. However, to avoid confusion, we refer to the solution only as Buchdahl solution for its simple form. On the other hand, the WH solution of Eq.(14) is threaded by a phantom scalar field, or \textit{imaginary} $\phi $, so that the stress tensor is negative, $\phi_{,\rho }\phi _{,\sigma }<0$. To our knowledge, the WH solution was simultaneously discovered by Ellis \cite{Ellis:1973} and Bronnikov \cite{Bronnikov:1973} in 1973. (These solutions are sometimes called the "Anti-Fisher" solution for change of sign before the stress term.) That is why, to be historically correct, we call them EBWH (Sec.2c).}:
\begin{eqnarray}
d\tau_{\text{{BNS}}}^{2} &=& -\left( 1+\frac{m}{2r}\right) ^{-2\gamma}\left(1-\frac{m}{2r}\right)^{2\gamma}dt^{2} \nonumber \\
&& + \left(1-\frac{m}{2r}\right)^{2(1-\gamma)}\left(1+\frac{m}{2r}\right)^{2(1+\gamma)} \nonumber \\
&& \times[dr^{2}+ r^{2}(d\theta^{2}+\sin ^{2}\theta d\psi ^{2})], \\
\phi_{\text{BNS}} &=& \left[ \left( \frac{\omega +3/2}{\kappa }\right) \left(\frac{C^{2}}{\lambda ^{2}}\right) \right] ^{1/2}\ln \left[ \frac{1-\frac{m}{2r}}{1+\frac{m}{2r}}\right], \\
\gamma &=& \frac{1}{\lambda }\left( 1+\frac{C}{2}\right).
\end{eqnarray}
The expression for $\lambda^{2}$, of course, continues to be the same as
Eq. (8), and using this, we can rewrite Eq. (17) as%
\begin{equation}
\phi _{\text{BNS}}=\left[ 2(1-\gamma ^{2})\right] ^{1/2}\ln \left[ \frac{1-%
\frac{m}{2r}}{1+\frac{m}{2r}}\right] .
\end{equation}%
The solution is valid for $r>m/2$ and the coordinate frame Ricci scalar $%
\widetilde{\mathbf{R}}$ is given by%
\begin{equation}
\widetilde{\mathbf{R}}=\frac{2m^{2}r^{4}(1-\gamma ^{2})}{\left( r-m/2\right)
^{2(2-\gamma )}\left( r+m/2\right) ^{2(2+\gamma )}}.
\end{equation}

$\bullet $ For $0<\gamma <1$ in Eq.(19), $\phi $ is real so that $\phi_{,\rho}\phi _{,\sigma }>0$ and $\widetilde{\mathbf{R}}>0$ from Eq.(20). There appears a singularity at $r_{s}=m/2$, where all the curvature scalars diverge as inherited from its predecessor JF Brans I solution (4). We call this singularity, which is not covered by a horizon, the \textit{BNS.} Note that the BNS is precisely the famous JNW NS written only in a different form \cite{Svitek:2016}.

$\bullet $ For $1<\gamma <\infty $ in Eq.(19), $\phi $ becomes imaginary so that $\phi _{,\rho }\phi _{,\sigma }<0$ and $\widetilde{\mathbf{R}}<0$ from Eq.(20) characteristic of WHs threaded by ghost matter, which we call \textit{EBWH}.

$\bullet $ For $\gamma =1$ in Eq.(19), one has $\phi _{,\rho}\phi _{,\sigma}=0$, and $\widetilde{\mathbf{R}}=0$ from Eq.(20) recovering \textit{SBH} in isotropic coordinates.

The ADM mass $M$ for the Buchdahl metric (16) is given by%
\begin{equation}
M_{\text{ADM}}=M=m\gamma .
\end{equation}%
This mass coincides also with the Keplerian mass $M_{\text{Kepler}}$ as may be found by the PPN expansion of the metric components (16), as well as with the ADM mass, $M_{\text{ADM}}$, i.e., for the Buchdahl NS,%
\begin{equation}
M_{\text{Kepler}}=M_{\text{ADM}}\Rightarrow r_{s}=\frac{M}{2\gamma}.
\end{equation}%
Under a coordinate transformation
\begin{equation}
\rho =r\left( 1+\frac{m}{2r}\right) ^{2},
\end{equation}%
one has%
\begin{equation}
1-\frac{2m}{\rho} = \left( \frac{1-\frac{m}{2r}}{1+\frac{m}{2r}}\right)^{2}
\end{equation}%
and so Buchdahl solution (16,19) transform to another form of JNW NS \cite{Virbhadra:1997}
\begin{eqnarray}
d\tau_{\text{{JNW}}}^{2} &=& -\left(1-\frac{2m}{\rho}\right)^{\gamma}dt^{2} + \left(1-\frac{2m}{\rho}\right)^{-\gamma}d\rho^{2}  \nonumber \\
&& +\rho^{2}\left(1-\frac{2m}{\rho}\right)^{(1-\gamma)}[d\theta^{2}+\sin^{2}\theta d\psi^{2}], \\
\phi_{\text{JNW}} &=& \left[2(1-\gamma^{2})\right] ^{1/2}\ln\left(1-\frac{2m}{\rho}\right) .
\end{eqnarray}%
Here also, when $\gamma =1,$ we recover SBH in standard coordinates with the NS now appearing at $\rho =2m$. We shall however calculate accretion properties using the isotropic form, which is the Buchdahl metric (16,19).

\subsection{JF Brans I $\longleftrightarrow$ JF Brans II}
To obtain the non-singular solution, we need to first remove the above mentioned singularity from the \textit{Brans I} solution (4-8). We can do it by the following operations on it:%
\begin{eqnarray}
r&\rightarrow& \frac{1}{r^{\prime }},\quad m\rightarrow \frac{2i}{m},\quad
\lambda \rightarrow -i\Lambda ,\quad \alpha _{0}\rightarrow \epsilon_{0}, \nonumber \\
\beta _{0}&\rightarrow& \delta _{0}+2\ln {\frac{m}{2}},
\end{eqnarray}%
where $m$ and $\Lambda $ are real. Using the identity%
\begin{equation}
\tan ^{-1}(x)\equiv \frac{i}{2}\ln \left( \frac{1-ix}{1+ix}\right) ,
\end{equation}%
we arrive at the\textit{\ JF Brans II} metric and the scalar field as follows%
\begin{equation}
d\tau _{\text{{Brans II}}}^{2}=-e^{2\alpha (r^{\prime })}dt^{2}+e^{2\beta (r^{\prime })}[dr^{\prime 2}+r^{\prime 2}(d\theta ^{2}+\sin ^{2}\theta d\psi^{2})],
\end{equation}%
where
\begin{eqnarray}
\alpha (r^{\prime}) &=& \epsilon _{0}+\frac{2}{\Lambda }\tan ^{-1}\left( \frac{2r^{\prime }}{m}\right), \\
\beta (r^{\prime}) &=& \delta _{0}-\frac{2(C+1)}{\Lambda }\tan ^{-1}\left(\frac{2r^{\prime }}{m}\right) \nonumber \\
&&-\ln \left( \frac{4r^{\prime 2}}{4r^{\prime 2}+m^{2}}\right), \\
\varphi_{\text{Brans II}}(r^{\prime}) &=& \varphi _{0}\exp \left[ \frac{2C}{\Lambda}\tan^{-1}\left(\frac{2r^{\prime }}{m}\right)\right], \\
\Lambda^{2} &\equiv& C\left( 1-\frac{\omega C}{2}\right) -(C+1)^{2}>0.
\end{eqnarray}%
Asymptotic flatness at $r^{\prime} = \infty $ requires that
\begin{equation}
\epsilon_{0} = -\frac{\pi}{\Lambda}, \quad \delta_{0} = \frac{\pi (C+1)}{\Lambda}.
\end{equation}

The story is now quite different - there is a drastic change. The above solution (29-33) has been listed by Brans \cite{Brans:1962b}, which we call Brans II solution, but we see that the Brans I and II are \textit{not} independent $-$ one can be derived from the other by the transformations in (27). However, though not independent, Brans I and II solutions are by no means equivalent as the former is singular, while the latter is a \textit{regular WH} \cite{Bhattacharya:2009}. It should be mentioned that the WH interpretation does not take away the WH's ability to explain the weak field tests of usual solar or stellar gravity in the Keperian positive mass ($\frac{2m}{\Lambda }$) on one side of the WH \cite{Izmailov:2020}. Here we have shown the passage, Brans I $\rightarrow$ Brans II, while the reverse passage (Brans II $\rightarrow $ Brans I) was shown by Bhadra and Sarkar \cite{Bhadra:2005}.

\subsection{JF Brans II $\longleftrightarrow$ EBWH}
JF Brans II solution was interpreted as a traversible twice asymptotically flat WH in \cite{Bhattacharya:2011}. We may go over to the EF and see that the same interpretation of WH still holds \cite{Bhattacharya:2011}. After redefining the constants in JF Brans II (29-33) as
$$\frac{2C}{\Lambda}\rightarrow 4\delta,\quad \frac{2(C+2)}{\Lambda}\rightarrow 4\gamma,\quad -\frac{2\pi}{\Lambda}\rightarrow \epsilon,$$
\begin{equation}
\frac{2\pi (1+C)}{\Lambda}\rightarrow \zeta,
\end{equation}%
and going over to the EF\ via using (11,12), we obtain a solution that we claim to be just the everywhere regular EF EBWH describing the positive mass mouth. This is given by
\begin{equation}
d\tau _{\text{EB{WH}}}^{2}=-P(r^{\prime })dt^{2}+Q(r^{\prime })[dr^{\prime 2}+r^{\prime 2}(d\theta ^{2}+\sin ^{2}\theta d\psi ^{2})],
\end{equation}%
where%
\begin{eqnarray}
P(r^{\prime}) &=& \exp \left[ 2\epsilon +4\gamma \tan ^{-1}(2r^{\prime }/m)\right], \\
Q(r^{\prime}) &=& \left( 1+\frac{m^{2}}{4r^{\prime 2}}\right) ^{2}\exp \left[2\zeta -4\gamma \tan ^{-1}(2r^{\prime }/m)\right],
\end{eqnarray}%
\begin{equation}
\phi_{\text{EBWH}}(r^{\prime}) =4\delta \tan^{-1}(2r^{\prime}/m),\quad 2\delta ^{2}=1+\gamma ^{2},
\end{equation}%
where the coordinate patch covers only one mouth or half-patch, $0<r^{\prime}<\infty$. Asymptotic flatness requires that $\epsilon =-\pi \gamma $ and $\zeta =\pi \gamma $. \textit{We shall call the metric (36-39) simply EBWH without the prefix EF and use it for calculation since accretion is taking place only around the positive mass mouth. }The constraint equation $2\delta^{2}=1+\gamma ^{2}$ comes from the EF field equations when the solution is put into them. This EF EBWH has a throat radius
\begin{equation}
r_{\text{th}}^{\prime }=\frac{M}{2\gamma }\left[ \gamma +\sqrt{1+\gamma ^{2}}\right] .
\end{equation}%

The passage from $d\tau _{\text{EBWH}}^{2}$ to the SBH\ $d\tau _{\text{SBH}}^{2}$ of mass $M$ in isotropic coordinates is possible under a combination of inversion and Wick rotation
\begin{equation}
r\rightarrow \frac{m^{2}}{4\rho },\text{ }\gamma =i,
\end{equation}%
and use of the identity
\begin{equation}
\tanh ^{-1}(x)\equiv \frac{1}{2}\ln \left( \frac{1+x}{1-x}\right) .
\end{equation}%
These reduce $d\tau _{\text{EBWH}}^{2}$ to
\begin{eqnarray}
d\tau_{\text{SBH}}^{2} &=& -\left(\frac{1-\frac{M}{2\rho}}{1+\frac{M}{2\rho}}\right)^{2}dt^{2}+\left(1+\frac{M}{2\rho}\right)^{4}\left[ d\rho^{2}\right. \nonumber \\
&&\left.+\rho ^{2}\left( d\theta ^{2}+\sin ^{2}\theta d\varphi ^{2}\right)\right].
\end{eqnarray}

Let us extend the coordinates to cover\textit{\ }both mouths by transforming
the radial variable $r^{\prime }\rightarrow \ell $ by
\begin{equation}
\ell =r^{\prime }-\frac{m^{2}}{4r^{\prime }}.
\end{equation}%
so that the solution (36-39) goes over into%
\begin{eqnarray}
d\tau_{\text{{EBWH}}}^{2} &=& -F(\ell )dt^{2}+F^{-1}(\ell )[d\ell ^{2}+(\ell^{2}+m^{2}) \nonumber \\
&& \times (d\theta ^{2}+\sin ^{2}\theta d\psi^{2})], \\
F(\ell) &=& \exp\left[ -2\pi \gamma +4\gamma \tan ^{-1}\left( \frac{\ell +\sqrt{\ell ^{2}+m^{2}}}{m}\right)\right],
\end{eqnarray}
\begin{eqnarray}
\phi_{\text{EBWH}}(\ell ) &=& 4\delta \tan ^{-1}\left( \frac{\ell +\sqrt{\ell^{2}+m^{2}}}{m}\right), \\
2\delta^{2} &=& 1+\gamma^{2},
\end{eqnarray}%
where $\ell $ now covers the two-sided domain, $-\infty <\ell <+\infty$. The throat now appears at $\ell _{\text{th}}=m\gamma $. Now use the identity%
$$\tan ^{-1}\left( \frac{\ell }{m}\right) \equiv 2\tan ^{-1}\left( \frac{\ell +\sqrt{\ell ^{2}+m^{2}}}{m}\right) -\frac{\pi }{2}.$$
Then the metric (45-48) yields precisely the familiar form of the solution obtained independently by Ellis \cite{Ellis:1973} and Bronnikov \cite{Bronnikov:1973}, hence the terminology EBWH:
\begin{eqnarray}
F_{\text{EBWH}}(\ell) &=& \exp \left[ -\pi \gamma +2\gamma \tan ^{-1}\left(\frac{\ell }{m}\right)\right], \\
\phi _{\text{EBWH}}(\ell) &=& 2\delta \left[ \frac{\pi }{2}+\tan ^{-1}\left(\frac{\ell }{m}\right)\right],  \\
2\delta ^{2} &=& 1+\gamma^{2}.
\end{eqnarray}

\subsection{BNS $\longleftrightarrow$ EBWH}
Using the coordinate transformation
\begin{equation}
\ell =r+\frac{m^{2}}{4r},
\end{equation}%
the BNS(16,19) can be expressed as

\begin{eqnarray}
d\tau _{\text{{BNS}}}^{2} &=&-f(\ell )dt^{2}+\frac{1}{f(\ell )}\left[ d\ell^{2}+(\ell ^{2}-m^{2})\right. \nonumber \\
&&\left.\times \left(d\theta ^{2}+\sin ^{2}\theta d\psi^{2}\right)\right] , \\
f(\ell ) &=&\left( \frac{\ell -m}{\ell +m}\right) ^{\gamma }, \\
\phi _{\text{BNS}}(\ell ) &=&\sqrt{2(1-\gamma ^{2})}\ln \left[ \frac{\ell -m}{\ell +m}\right].
\end{eqnarray}%
In this form, it is exactly the Ellis I solution \cite{Buchdahl:1959} that has been discussed also by Bronnikov and Shikin \cite{Bronnikov:1988}. The singularity has now been shifted to $\ell =m$.

To remove the singularity, we analytically continue the BNS by means of Wick rotation of the parameters while maintaining the positivity of the ADM mass $M=m\gamma $. We choose
\begin{equation}
m\rightarrow -im,\quad \gamma \rightarrow i\gamma .
\end{equation}%
This results in just the EBWH with the metric%
\begin{eqnarray}
d\tau_{\text{{EBWH}}}^{2} &=& -f(\ell )dt^{2}+\frac{1}{f(\ell )}\left[d\ell^{2}+(\ell ^{2}+m^{2}) \right. \nonumber \\
&&\times\left.\left( d\theta ^{2}+\sin ^{2}\theta d\psi ^{2}\right)\right], \\
f(\ell ) &=& \exp \left[ -2\gamma \text{cot}^{-1}\left( \frac{\ell }{m}\right)\right], \\
\phi_{\text{EBWH}}(\ell) &=& \left[ \sqrt{2\left( 1+\gamma ^{2}\right) }\right]\text{cot}^{-1}\left( \frac{\ell }{m}\right),
\end{eqnarray}
which can be expressed in the original form (49-51) by using the identities
\begin{eqnarray}
\text{cot}^{-1}(x)+\tan ^{-1}(x) &=&+\frac{\pi }{2},\quad x>0 \\
&=&-\frac{\pi }{2},\quad x<0.
\end{eqnarray}%
To recapitulate, the BNS (53-55) reduces to SBH for $\gamma =1$, NS for $\gamma <1$ and for $\gamma >1$, a non-traversable WH due to the presence of singularity at $\ell =m$. The NS can be regularized by the complex transformations (56) to obtain the EBWH, which also leads to SBH but for $\gamma =i$.

\section{Massless corollaries}
\label{sec:3}
These are rather curious cases.

(i)\ When $m=0$ in the EBWH\ (36-39), the spacetime becomes flat, as
expected. But when $m\neq 0$ but $\gamma =0$, we obtain what is known as a
symmetric \textit{massless EBWH}, because the ADM\ mass is\textit{\ }%
individually zero on each side. However, the integrated scalar field energy
is of equal and opposite signs on two sides, so they also add to zero. This
means the individual sides exhibit gravitational action on light \cite{Izmailov:2019b}. The
massless EBWH does not accrete matter since $r_{\text{ms}}^{\text{EBWH}%
}\rightarrow \infty $ (see Sec.5). However, it can bend light rays \textit{%
around} its throat and can thus be detected by observation of gravitational
lensing. This massless object has been considered as a candidate for a
possible halo obect in our galaxy \cite{Abe:2010,Lukmanova:2016}.

The metric (57-59) at $\gamma =0$ reduces to%
\begin{eqnarray}
d\tau _{\text{{EBWH,}}M=0}^{2} &=&-dt^{2}+d\ell ^{2}+(\ell ^{2}+m^{2}) \nonumber \\
&&\times\left(d\theta ^{2}+\sin^{2}\theta d\psi ^{2}\right) , \\
\phi _{\text{EBWH,}M=0} &=&\sqrt{2}\left[ \frac{\pi }{2}+\tan ^{-1}\left(\frac{\ell }{m}\right) \right] .
\end{eqnarray}%
This metric can mimic a BH in producing gravitational waves as analyzed in
\cite{Konoplya:2016}. The metric (62-63) represents a geodesically complete manifold on
which $\ell \in (-\infty ,\infty )$ and the throat appears at $\ell _{\text{%
th}}=0$. Light bends towards the source in the sector $0<\ell <\infty $
(attractive gravity) and bends away from the source in the sector $-\infty
<\ell <0$ (repulsive gravity). Using the transformation%
\begin{equation}
R=\sqrt{\ell ^{2}+m^{2}},
\end{equation}%
the metric (62,63) in standard coordinates becomes%
\begin{eqnarray}
d\tau _{\text{{EBWH,}}M=0}^{2} &=&-dt^{2}+\frac{dR^{2}}{1-\frac{m^{2}}{R^{2}}}+R^{2} \nonumber \\
&&\times\left(d\theta ^{2}+\sin ^{2}\theta d\psi^{2}\right) , \\
\phi _{\text{EBWH,}M=0} &=&\sqrt{2}\left[ \frac{\pi }{2}+\tan ^{-1}\left(\frac{\sqrt{R^{2}-m^{2}}}{m}\right) \right]
\end{eqnarray}%
and the throat shifts to $R_{\text{th}}^{\text{EBWH,}M=0}=m.$ The two-way
light bending angle is (see, for details, \cite{Bhattacharya:2010,Tsukamoto:2016})
\begin{equation}
\alpha _{M=0}^{\text{EBWH}}(R_{0})=\frac{\pi m^{2}}{4R_{0}^{2}}+\frac{9\pi
m^{4}}{64R_{0}^{4}}+...
\end{equation}%
where $R_{0}>m$ is the closest approach distance and shows deflection
\textit{towards} the throat.

(ii) The BNS at $\gamma =0$ has the metric ($m/2<r<\infty $):%
\begin{eqnarray}
d\tau _{\text{{BNS,}}M=0}^{2} &=&-dt^{2}+\left( 1-\frac{m^{2}}{4r^{2}}\right)^{2}[dr^{2}+r^{2} \nonumber \\
&&\times(d\theta ^{2}+\sin ^{2}\theta d\psi ^{2})], \\
\phi _{\text{BNS,}M=0} &=&\sqrt{2}\ln \left[ \frac{1-\frac{m}{2r}}{1+\frac{m}{2r}}\right] \approx -\frac{m}{r},
\end{eqnarray}%
and the ADM mass $M=m\gamma =0$ ($m\neq 0$), while the \textit{massless
NS} at $r=m/2$ is made purely of the scalar charge $m$ (it is
still called naked since curvature scalars diverge at $r=m/2$ and there is
no horizon). This metric does not accrete matter as $r_{\text{ms}}^{\text{BNS%
}}$ is \textit{imaginary} at $\gamma =0$ but it can nonetheless cause light
bending which, following the method in \cite{Bhattacharya:2010}, is
\begin{equation}
\alpha _{M=0}^{\text{BNS}}(R_{0})=-\frac{\pi m^{2}}{4R_{0}^{2}}+\frac{9\pi
m^{4}}{64R_{0}^{4}}+...
\end{equation}%
so the leading order term shows the deflection \textit{away} from the
singularity at $r=m/2$ and is exactly negative to that of EBWH. This is in
stark qualitative contrast between two massless objects having implications
for gravitational lensing observables, which will be examined in detail
elsewhere.

\section{Thin accretion disk formulas}
\label{sec:4}
Below we briefly describe the developments in Harko \textit{et al.} \cite{Harko:2009b}. The accretion disc is formed by particles moving in circular orbits around a compact object, with the geodesics determined by the space-time geometry around the object, be it a WH or BH. For a static and spherically symmetric geometry the metric is given in a general form by

\begin{equation}
d\tau ^{2}=g_{tt}dt^{2}+g_{rr}dr^{2}+g_{\theta \theta }d\theta^{2}+g_{\varphi \varphi }d\varphi ^{2}.
\end{equation}%
At and around the equator, the metric functions $g_{tt}$, $g_{rr}$, $g_{\theta\theta}$ and $g_{\phi\phi}$ only depend on the radial coordinate $r$, and the thinness of the disk is defined by $\left\vert \theta -\pi /2\right\vert \ll 1.$

\subsection{Kinematic formulas}

These are the angular velocity $\Omega $, the specific energy $\widetilde{E}$%
, and the specific angular momentum $\widetilde{L}$ of particles moving in
circular orbits in a static and spherically symmetric geometry and are given
by:
\begin{equation}
\frac{dt}{d\tau}=\frac{\widetilde{E}}{-g_{tt}},\frac{d\varphi }{d\tau} = \frac{\widetilde{L}}{g_{\varphi \varphi }}
\end{equation}%
and
\begin{equation}
g_{rr}\left( \frac{dr}{d\tau }\right) ^{2}=-1+\frac{\widetilde{E}^{2}g_{\varphi \varphi }+\widetilde{L}^{2}g_{tt}}{-g_{tt}g_{\varphi\varphi}}.
\end{equation}%
The last equation provides an effective potential term
\begin{equation}
V_{\text{eff}}\left(r\right) = -1+\frac{\widetilde{E}^{2}g_{\varphi\varphi} + \widetilde{L}^{2}g_{tt}}{-g_{tt}g_{\varphi\varphi}}.
\end{equation}%
Existence of circular orbits in the equatorial plane demands that $V_{\text{%
eff}}\left( r\right) =0$ and $V_{\text{eff},r}\left( r\right) =0$, where the
comma in the subscript denotes a derivative with respect to the radial
coordinate $r$. These conditions allow us to write
\begin{eqnarray}
\widetilde{E}&=&-\frac{g_{tt}}{\sqrt{-g_{tt}-g_{\varphi \varphi }\Omega ^{2}}}, \quad
\widetilde{L}=\frac{g_{\varphi \varphi }\Omega }{\sqrt{-g_{tt}-g_{\varphi\varphi }\Omega ^{2}}}, \nonumber \\
\Omega &=& \frac{d\varphi }{dt}=\sqrt{\frac{-g_{tt,r}}{g_{\varphi \varphi ,r}}}.
\end{eqnarray}

We assume thin accretion disk with height $H$ much smaller than the
characteristic radius $R$ of the disk, $H\ll R$. The thin disk is assumed to
be in hydrodynamical equilibrium stabilizing its vertical size, with the
pressure and vertical entropy gradient being negligible in the disk. The
efficient cooling via the radiation over the disk surface is assumed
preventing the disk from collecting the heat generated by stresses and
dynamical friction. The thin disk has an inner edge defined by the \textit{%
marginally stable} orbit or innermost stable circular orbit (ISCO) radius $%
r_{\text{ms}}$ defined by the solution of $\left. d^{2}V_{\text{{eff}}%
}/dr^{2}\right\vert _{r=r_{\text{{ms}}}}=0$, while the orbits at higher
radii are Keplerian. In steady-state accretion disk models, the mass
accretion rate $\dot{M}_{0}$ is assumed to be a constant and the physical
quantities describing the orbiting matter are averaged over a characteristic
time scale, e.g., the total period of the orbits over the azimuthal angle $%
\Delta \varphi =2\pi $ , and over the height $H$ \cite{Shakura:1973,Novikov:1973,Page:1974}.

In the above steady-state thin disk model, the orbiting particles have $%
\Omega$, $\widetilde{E}$ and $\widetilde{L}$ that depend only on the radii
of the orbits. Accreting particles orbiting with the four-velocity $u^{\mu
} $ form a disk of an averaged surface density $\Sigma $. Page and Thorne
\cite{Page:1974}, using the rest mass conservation law, showed that the time averaged
rate of rest mass accretion $dM_{0}/dt$ is independent of the radius: $\dot{%
M_{0}}\equiv dM_{0}/dt=-2\pi ru^{r}\Sigma =$ \textmd{const}. (Here $u^{r}$
is the radial component of the four-velocity). We omit other technical
details of the model (for which, see \cite{Page:1974}) but quote only the relevant
formulas below.

\subsection{Emissivity formulas}

These consist of the flux $F(r)$, temperature $T(r)$ and the luminosity
of the radiant energy $L\left(\nu\right)$ over the disk that can be
expressed in terms of $\Omega$, $\widetilde{E}$ and $\widetilde{L}$ of the
compact sphere \cite{Shakura:1973,Novikov:1973,Page:1974}
\begin{equation}
F\left( r\right) =-\frac{\dot{M}_{0}}{4\pi \sqrt{-g}}\frac{\Omega _{,r}}{%
\left( \widetilde{E}-\Omega \widetilde{L}\right) ^{2}}\int_{r_{\text{ms}%
}}^{r}\left( \widetilde{E}-\Omega \widetilde{L}\right) \widetilde{L}_{,r}dr.
\end{equation}%
The accreting matter in the steady-state thin disk model is supposed to be
in thermodynamical equilibrium. Therefore the radiation flux emitted by the
disk surface will follow Stefan-Boltzmann law:%
\begin{equation}
F\left( r\right) =\sigma T^{4}\left( r\right) ,
\end{equation}%
where $\sigma $ is the Stefan-Boltzmann constant. The observed luminosity $%
L\left( \nu \right)$ has a redshifted black body spectrum \cite{Torres:2002}
\begin{equation}
L\left( \nu \right) =4\pi \text{d}^{2}I(\nu )=\frac{8\pi h\cos {j}}{c^{2}}%
\int_{r_{\text{ms}}}^{r_{\text{f}}}\int_{0}^{2\pi }\frac{\nu
_{e}^{3}rdrd\varphi }{\exp \left[ \frac{h\nu _{e}}{k_{B}T}\right] -1}.
\end{equation}%
Here $h$ is Planck's constant, $I\left( \nu \right) $ is the Planck
distribution function, $k_{B}$ is the Boltzmann constant, $\nu _{e}$ is the
emission frequency, d is the distance to the source, $j$ is the disk
inclination angle perpendicular to the line of sight, and $r_{\text{ms}}$
and $r_{\text{f}}$ indicate the position of the innermost and outermost edge
of the disk, respectively. We take $r_{\text{f}}\rightarrow \infty $, since
we expect that the flux over the disk surface vanishes at $r\rightarrow
\infty $ for any kind of general relativistic compact object described by
asymptotically flat geometry. We take $j=0^{\circ }$ so that the disk is
face-on. The observed photons are redshifted to the frequency $\nu $ related
to the emission frequency $\nu_{e}$ relatied by
\begin{equation}
\nu _{e}=(1+z)\nu,
\end{equation}%
where the red-shift factor is given by \cite{Harko:2008}
\begin{equation}
1+z=\frac{1+\Omega r\sin {\varphi }\sin {j}}{\sqrt{-g_{tt}-\Omega^{2}g_{\varphi \varphi }}}.
\end{equation}

Following Page-Thorne model, an important characteristic of the accretion
disk is the \textit{efficiency }$\epsilon $, which quantifies the ability by
which the accreting body converts particle mass into radiation. It is
measured at infinity and is defined as the ratio between the rate of energy
of the photons emitted from the disk surface and the rate with which
mass-energy is transported to the central accreting body. If all photons
reach asymptotic infinity, the efficiency is given by the specific energy $%
\widetilde{E}$ of accreting particles measured at $r=r_{\text{ms}}$ such
that \cite{Thorne:1974}

\begin{equation}
\epsilon =1-\widetilde{E}_{\text{ms}}.
\end{equation}%
We shall use the above formulas to numerically compute the kinematic and emissivity properties.

\section{Analyses of accretion profiles: BNS, EBWH and SBH}
\label{sec:5}
We shall henceforth be considering Eqs.(16-19) for BNS, Eqs.(36-39) for EBWH and Eq.(43) for SBH for the ensuing calculations. Note that we assume a toy model for the mass of the accreting central object to be $M=15M_{\odot}$ with the accretion rate $\dot{M}_{0}=10^{18}$ gm.sec$^{-1}$ for illustration.

\subsection{Behavior of kinematic profiles}

For the reality of the BNS, it is necessary that $0<\gamma <1$, while for
the EBWH, $1<\gamma <\infty $ , the intervals being mutually exclusive. The
SBH with a vanishing scalar field $\phi $ corresponds to $\gamma =1$ in BNS
and $\gamma =i$ in the EBWH. To derive the kinematic properties, consider
Eq.(74) for the\ generic effective potential $V_{\text{{eff}}}$. For the
solutions under consideration, we explicitly derive $r_{\text{{ms}}}$ in
isotropic coordinates as under [using $m=\frac{M}{\gamma }$ in (16) where $M$
is the $M_{\text{ADM}}$ defined in (21)]:
\begin{eqnarray}
r_{\text{ms}}^{\text{BNS}} &=& \frac{M}{2\gamma}\left[3\gamma + \sqrt{5\gamma^{2}-1} + \sqrt{2}\gamma\right. \nonumber \\
&&\times\left.\left(7- \frac{1}{\gamma^{2}}-\frac{3-15\gamma^2}{\gamma\sqrt{5\gamma^{2}-1}}\right)^{\frac{1}{2}}\right], \\
r_{\text{ms}}^{\text{EBWH}} &=& \frac{M}{2\gamma}\left[3\gamma + \sqrt{5\gamma^{2}+1} + \sqrt{2}\gamma\right. \nonumber \\
&&\times\left.\left(7 + \frac{1}{\gamma^{2}}+\frac{3+15\gamma^2}{\gamma\sqrt{5\gamma^{2}+1}}\right)^{\frac{1}{2}}\right], \\
r_{\text{ms}}^{\text{SBH}} &=& \frac{(5+2\sqrt{6})M}{2}=4.94949M,
\end{eqnarray}
which is a well known value. The kinematic formulas including the potential $%
V_{\text{eff}}\left( r\right) $ in Eq.(74) valid at an arbitrary radius $r$
are as under:
\begin{eqnarray}
\widetilde{E}^{\text{BNS}} &=& \left[ \frac{4r^{2}+M(M-4\gamma r)}{4r^{2}+M(M-8\gamma r)}\right] ^{\frac{1}{2}}\left( 1-\frac{M}{2r}\right)^{\gamma} \nonumber \\
&&\times \left(1+\frac{M}{2r}\right)^{-\gamma}, \\
\widetilde{L}^{\text{BNS}} &=& -\frac{1}{2}\left[ \frac{\gamma M(M^{2}-4r^{2})}{4r^{3}+Mr(M-8\gamma r)}\right] ^{\frac{1}{2}}\left(1-\frac{M}{2r}\right)
^{-\gamma} \nonumber \\
&&\times \left( 1+\frac{M}{2r}\right)^{\gamma},
\end{eqnarray}
\begin{eqnarray}
\widetilde{E}^{\text{EBWH}} &=& \left[\frac{-M^{2}+4Mr\gamma ^{2}+4r^{2}\gamma^{2}}{-M^{2}+8Mr\gamma ^{2}+4r^{2}\gamma ^{2}}\right] ^{\frac{1}{2}} \nonumber \\
&&\times \exp\left[-\pi\gamma+2\gamma \tan ^{-1}\left( \frac{2\gamma r}{M}\right)\right], \\
\widetilde{L}^{\text{EBWH}} &=& -\frac{1}{2}\left[ -\frac{M^{3}+4Mr^{2}\gamma^{2}}{M^{2}r+8Mr^{2}\gamma ^{2}-4r^{3}\gamma ^{2}}\right] ^{\frac{1}{2}} \nonumber \\
&&\times \exp\left[ \pi \gamma -2\gamma \tan ^{-1}\left( \frac{2\gamma r}{M}\right)\right].
\end{eqnarray}
The marginally stable radii in isotropic coordinates yield the ratio $r_{%
\text{ms}}^{\text{EBWH}}(\gamma \rightarrow \infty )/$ $r_{\text{ms}}^{\text{%
Sch}}=1.0579$. The demand for the reality of the expression for $r_{\text{ms}%
}^{\text{BNS}}$ in Eq.(82) further reduces the BNS range $0<\gamma <1$ to
the range $\frac{1}{\sqrt{5}}<\gamma <1$, which then corresponds to the $2.5<
$ $r_{\text{ms}}^{\text{BNS}}/M<4.949$ and the singular radii $0.5<$ $r_{%
\text{s}}^{\text{BNS}}/M<1.118$. These clearly show that $r_{\text{s}}^{%
\text{BNS}}<r_{\text{ms}}^{\text{BNS}}$. The EBWH range is $1<\gamma <\infty
$, which corresponds to minimally stable radii $5.495>$ $r_{\text{ms}}^{%
\text{EBWH}}/M>5.236$ and the throat radii $1.207>$ $r_{\text{th}}^{\text{%
EBWH}}/M>1$, indicating that $r_{\text{th}}^{\text{EBWH}}<$ $r_{\text{ms}}^{%
\text{EBWH}}$. The idea here is to show that marginally stable orbits for
BNS (or for EBWH) do not come close the NS since the allowed
radial intervals are \textit{disjoint}, not overlapping around a common
value $r_{\text{ms}}^{\text{BNS}}/M\sim r_{\text{s}}^{\text{BNS}}/M$, so it
is not possible that the $r_{\text{ms}}^{\text{BNS}}$ could approach $r_{%
\text{s}}^{\text{BNS}}$. As a result, we do not find arbitrarily large
increase in the accretion luminosity, unlike what happens in the NS [10] at $%
r_{\text{ms}}^{\text{NS}}\rightarrow r_{\text{s}}^{\text{NS}}=0$, the
central singularity. That is why we shall calculate the profiles only from $%
r_{\text{ms}}$ upwards as shown in the relevant figures. The distinction
between the two singularities probably signals the generic difference
between the BNS and NS in \cite{Joshi:2014} - the former singularity is removable while
that the latter is not. Further, in the former, $r=0$ is the central
singularity, whereas in the latter the center is a regular point.

Note once again that singularity appears at $r_{\text{s}}^{\text{BNS}}=\frac{%
M}{2\gamma }$, horizon at $r_{\text{s}}^{\text{SBH}}=\frac{M}{2}$ and the
throat at $r_{\text{th}}^{\text{EBWH}}=\frac{M}{2\gamma }\left[ \gamma +%
\sqrt{1+\gamma ^{2}}\right] .$ Eq.(85) yields $\widetilde{E}^{\text{BNS}%
}\rightarrow 0$ as $r_{\text{s}}^{\text{BNS}}\rightarrow \frac{M}{2\gamma }$%
, which in turn yields $\epsilon \rightarrow 1$ or $100\%$ efficiency in
exact accordance with the result obtained in \cite{Joshi:2014} about the efficiency of
their NS by integrating the spectral luminosity\footnote{%
We are thankful to the anonymous second referee for pointing out the result
in \cite{Joshi:2014}.}. However, according to Page-Thorne model, some illustrative values
of $\epsilon $ are: For SBH it is $0.0572$ or $5.72\%$ at $r_{\text{ms}}^{%
\text{SBH}}/M=4.949$, for EBWH it is $5.42\%$ at $r_{\text{ms}}^{\text{EBWH}%
}/M=5.303$ and for BNS it is $7.9\%$ at $r_{\text{ms}}^{\text{BNS}}/M=2.809$%
, as in Table 1. These results qualitatively accord with those in \cite{Kovacs:2010}.

\begin{table}[!ht]
\caption{The $r_{\text{ms}}$ and the efficiency $\epsilon $ for EBWH, BNS and SBH. Note that $\gamma =1$ in BNS and $\gamma =i$ in EBWH respectively yield SBH quantities, while $\gamma =0$ yields the values for massless BNS and EBWH respectively.}
\centering
\begin{tabular}{|c|c|c|c|c|}
\hline
$\gamma $ & \multicolumn{2}{|c|}{EBWH} & \multicolumn{2}{|c|}{BNS} \\
\cline{2-5}
& $r_{\text{ms}}$/$M$ & $\epsilon $ & $r_{\text{ms}}$/$M$ & $\epsilon $ \\
\hline
$0$ & $\infty $ & $-$ & imaginary & $-$ \\ \hline
$0.45$ & $-$ & $-$ & $2.809$ & $0.0790$ \\
$0.5$ & $-$ & $-$ & $3.732$ & $0.0693$ \\
$0.7$ & $-$ & $-$ & $4.609$ & $0.0603$ \\
$0.9$ & $-$ & $-$ & $4.877$ & $0.0578$ \\
$1$ & $-$ & $-$ & $4.949$ & $0.0572$ \\
$1.25$ & $5.404$ & $0.0534$ & $-$ & $-$ \\
$1.5$ & $5.354$ & $0.0538$ & $-$ & $-$ \\
$2$ & $5.303$ & $0.0542$ & $-$ & $-$ \\
$2.5$ & $5.279$ & $0.0544$ & $-$ & $-$ \\
$i$ & $4.949$ & $0.0572$ & $-$ & $-$ \\ \hline
\end{tabular}
\end{table}

The kinematic profiles are presented in the Figs.2,3,4. The effective
potential $V_{\text{eff}}(r)$ for the three objects described by a EBWH and
BNS are shown in Fig.2. The specific angular momentum of the orbiting
particle is chosen to be $\tilde{L}=4M$. From an examination of the plots
for EBWH (\textit{top panel, left hand}), it turns out that the potential
barrier peak is \textit{highest} at the SBH limit ($\gamma =i$), while it is
depressing with the decrease in the value of $\gamma $. For BNS (\textit{top
panel, right hand}), it is seen that the picture is just the opposite. The
barrier peak is \textit{lowest} at the SBH limit ($\gamma =1$), while the
barrier is rising indefinitely with the decrease in the value of $\gamma $,
finally diverging at $r_{s}=\frac{M}{2\gamma}$ indicative of NS. A strict comparison between the accretion profiles of EBWH, BNS
and SBH is impossible since they are defined within mutually exclusive
intervals of $\gamma$. However, the profiles of EBWH and BNS may be
individually compared with those of SBH, which is a unique limit having a
vanishing scalar field $\phi$. This comparison in potential profiles of BNS
and EBWH are shown for fixed relevant values of $\gamma $ and compared with
those of SBH (\textit{bottom panel}). The BNS profiles show maxima at a
height much higher than those for the other objects, which characterizes NS.

The specific energy $\tilde{E}(r)$ of the orbiting particles as a function
of the radial coordinate $r$ (in cm) for the three objects are displayed in
Fig.3. The profiles of SBH appear between the higher profile of BNS and
lower ones due to EBWH (\textit{bottom panel}). The behavior of the specific
angular momentum $\tilde{L}(r)$ of orbiting particles are shown in Fig.4.
The overall pattern for EBWH is that $\tilde{L}(r)$ decreases from above to
a minimum and then increases again (\textit{top panel, left hand}). For the
BNS, similar behavior is exhibited only for $\gamma \leq 0.5$, whereas for
higher values of $\gamma $, $\tilde{L}(r)$ increases but eventually
approaches the previous minimum but from below (\textit{top panel, right hand%
}). Patterns similar to that for potential are also seen for relevant values
of $\gamma $ (\textit{bottom panel}).

\begin{figure*}[!ht]
  \centerline{\includegraphics[scale=1.8]{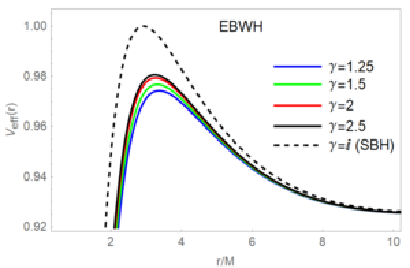} \includegraphics[scale=1.8]{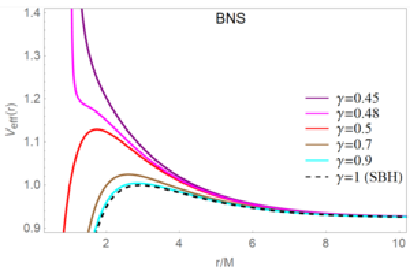}}
  \centerline{\includegraphics[scale=1.8]{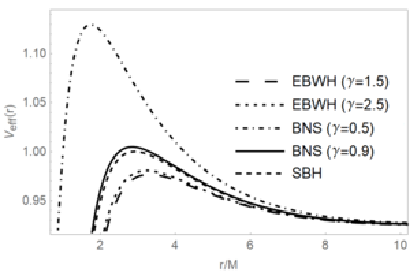}}
  \caption{The effective potential $V_{\text{eff}}$ for EBWH ($\protect\gamma >1$), BNS ($\protect\gamma <1$) and SBH ($\protect\gamma =-i$ from EBWH, and $\protect\gamma =1$ from BNS).}
  \label{Veff}
\end{figure*}

\begin{figure*}[!ht]
  \centerline{\includegraphics[scale=1.8]{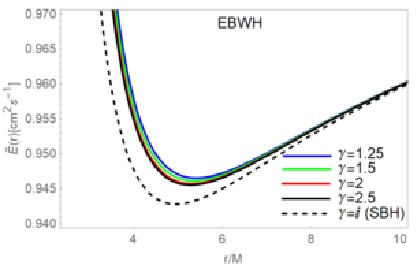} \includegraphics[scale=1.8]{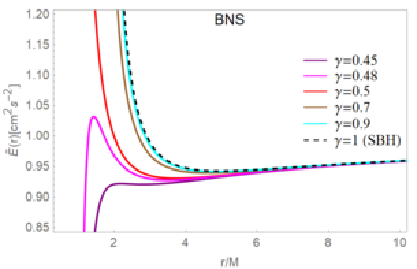}}
  \centerline{\includegraphics[scale=1.8]{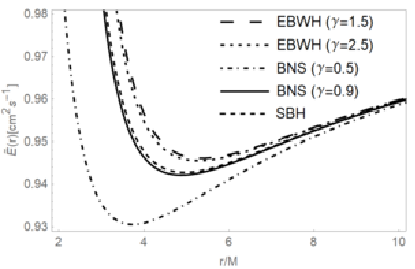}}
  \caption{The specific energy $\widetilde{E}(r)$ for EBWH, BNS and SBH for different values of $\protect\gamma$.}
  \label{Energy}
\end{figure*}

\begin{figure*}[!ht]
  \centerline{\includegraphics[scale=1.8]{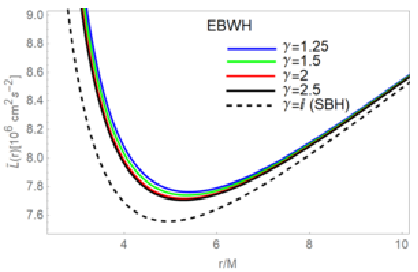} \includegraphics[scale=1.8]{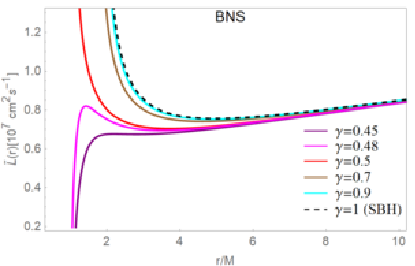}}
  \centerline{\includegraphics[scale=1.8]{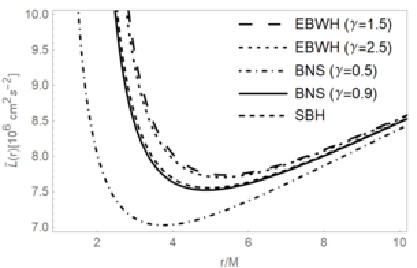}}
  \caption{The specific angular momentum $\widetilde{L}(r)$ for EBWH, BNS and SBH for different values of $\protect\gamma$.}
  \label{AngMom}
\end{figure*}

\begin{figure*}[!ht]
  \centerline{\includegraphics[scale=1.8]{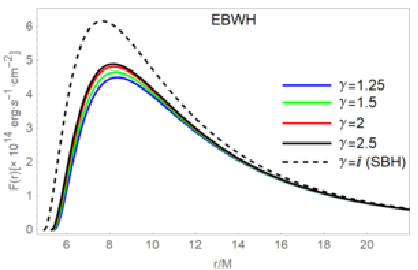} \includegraphics[scale=1.8]{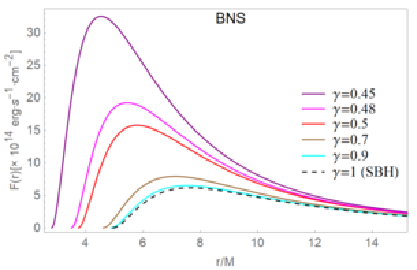}}
  \centerline{\includegraphics[scale=1.8]{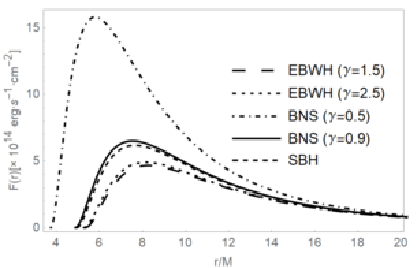}}
  \caption{The time averaged radiation flux $F(r)$ for EBWH, BNS and SBH for different values of $\protect\gamma $.}
  \label{Flux}
\end{figure*}

\begin{figure*}[!ht]
  \centerline{\includegraphics[scale=1.8]{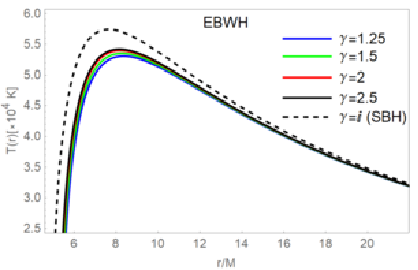} \includegraphics[scale=1.8]{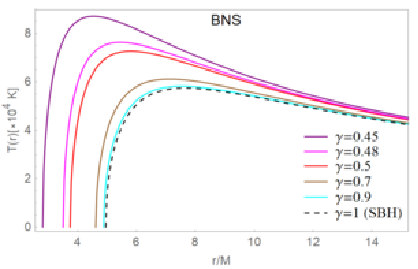}}
  \centerline{\includegraphics[scale=1.8]{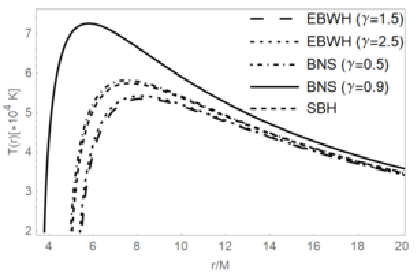}}
  \caption{Temperature distribution $T(r)$ of the accretion disk for EBWH, BNS and SBH for different values of $\protect\gamma $.}
  \label{Temp}
\end{figure*}

\begin{figure*}[!ht]
  \centerline{\includegraphics[scale=1.8]{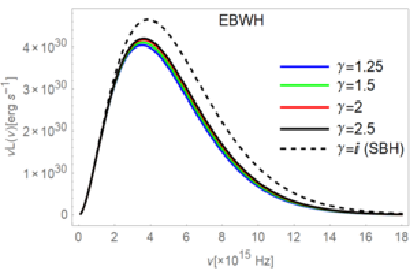} \includegraphics[scale=1.8]{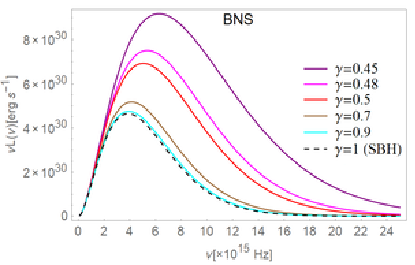}}
  \centerline{\includegraphics[scale=1.8]{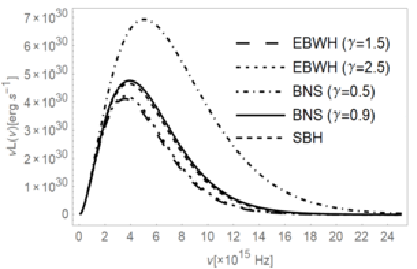}}
  \caption{The emission spectra $\protect\upsilon L(\protect\upsilon)$ of the accretion disk with inclination $i=0^{\circ }$\ for EBWH, BNS and SBH for different values of $\protect\gamma $.}
  \label{Lum}
\end{figure*}

\subsection{Behavior of emissivity profiles}

Using relevant values of $\gamma$, the Figs.5,6,7 show the emissivity
profiles, all of which nearly coincide at large $r$. Fig.5 shows the time
averaged flux $F(r)$ as a function of the radial coordinate $r$
radiated by the disk for EBWH (\textit{top panel, left hand}) shows that the
flux peaks highest for SBH ($\gamma =i$) than that of EBWH for different
real $\gamma$. This behavior is qualitatively just the opposite in the case
of BNS, where its maximum appears far above that of the SBH, $\sim 6$ times
higher (\textit{top panel, right hand}). Comparison of the\ radiation flux
profiles are also seen between BNS and EBWH for respective admissible values
of $\gamma$ (\textit{bottom panel}). It can be observed that the EBWH
profiles appear (\textit{top panel, left hand}) somewhere in between those
of BNS and SBH. Fig.6 shows temperature distribution $T(r)$ as a function of
the radial coordinate $r$ of the accretion disk for EBWH (\textit{%
top panel, left hand}), BNS (\textit{top panel, right hand}), and comparison
in the temperature distribution $T(r)$ is seen between BNS and EBWH for
respective admissible values of $\gamma $ (\textit{bottom panel). }The
qualitative features are the same as that of flux, as expected, since it is
connected to radiation flux by Stefan-Bolzmann law. Fig.7 shows the
luminosity spectra $\nu L(\nu )$ of the accretion disk for EBWH \textit{(top
panel, left hand}), BNS (\textit{top panel, right hand}). The luminosity
profiles of these objects are compared with that of SBH for relevant values
of $\gamma $ (\textit{bottom panel}). The overall patterns show that SBH
profile dominates over that of EBWH but the BNS profiles dominate over both.
Indeed, in the case of a NS, the $r_{ms}$ comes closer to the
central singularity, which leads to a considerable increase in the
luminosity of the disk though it does not lead to $100\%$ efficiency.

Joshi et al \cite{Joshi:2014} defined a differential luminosity reaching an observer at
infinity as
\begin{equation}
\frac{d\mathcal{L}_{\infty }}{d\ln {r}}=4\pi r\sqrt{-g}\widetilde{E}F,
\end{equation}%
where $\widetilde{E}$ is the specific energy and $F$ is the radiation flux
emitted by the disk. Figs.8,9 show differential luminosity for BNS, EBWH and
SBH, respectively. We obtain the same SBH\ Page-Thorne profile as obtained
in \cite{Joshi:2014}. The most significant difference between BNS and SBH (Fig.8) appears
near $r_{\text{ms}}$, where the differential luminosity of BNS is always
noticeably larger than SBH (but not infinitely large). In contrast,
differential luminosity of EBWH is always lower than that of SBH. However, a
very important observable - the Eddington luminosity at infinity\ shows
\textit{infinite} magnitude as the singular radius $r_{s}$ is approached,
which we discuss next.

\begin{figure}[!ht]
  \centerline{\includegraphics[scale=1.8]{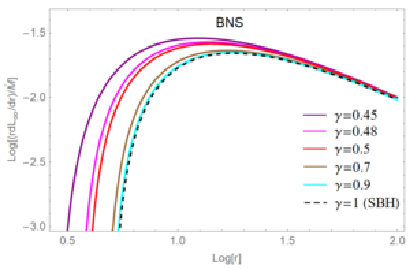}}
  \caption{Profile of the BNS differential luminosity, $\left( rd\mathcal{L}_{\infty }/dr\right) /\dot{M}_{0}$, for different $\protect\gamma $.}
  \label{DifLumBNS}
\end{figure}

\begin{figure}[!ht]
  \centerline{\includegraphics[scale=1.8]{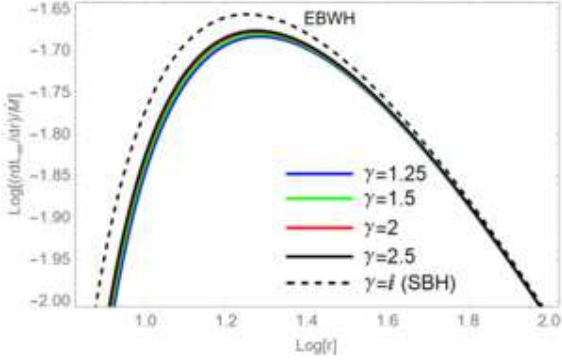}}
  \caption{Profile of the EBWH differential luminosity reaching an observer at infinity, $\left( rd\mathcal{L}_{\infty }/dr\right) /\dot{M}_{0}$, for different $\protect\gamma $.}
  \label{DifLumEBWH}
\end{figure}

\section{Eddington luminosity of the disk}
\label{sec:6}
The Eddington luminosity\footnote{%
We thank the anonymous first referee for drawing our attention to this
property of the disk.}, often called the Eddington limit, is the maximum
luminosity that an accreting disk around an astrophysical body can achieve,
when there is balance between the force of radiation acting outward and the
gravitational force acting inward establishing a hydrostatic equilibrium.
This luminosity for the case of a boson star was pointed out in \cite{Torres:2002}. The
hydrostatic equilibrium leads to Eddington luminosity given by $L_{\text{Edd}%
}=4\pi Mm_{p}/\sigma _{T}=1.3\times 10^{38}\left( M/M_{\odot }\right) $
erg/s, where $m_{p}$ is the proton mass, and $\sigma _{T}$ is the Thompson
cross section \cite{Bini:2009}. Since the accretion disk in general has a radius
dependent mass distribution, $M=M(r)$, the Eddington luminosity becomes a
radius dependent quantity as well, $L_{\text{Edd}}(r)\propto M(r)$. A
similar effect can also occur in the case of accretion disk around the NS and WHs considered in the present paper. Except SBH, our
solutions for BNS and EBWH are sourced by the scalar fields (Eqs.(19) and
(39) respectively) that can be described by a energy-momentum tensor
yielding a mass distribution $M(r)$ along the equatorial plane of the disk
\cite{Kovacs:2010}, given by $M(r)=4\pi \int_{r_{\bullet }}^{r}T_{0}^{\varphi
0}r^{2}dr=2\pi \int_{r_{\bullet }}^{r}g^{rr}\varphi _{,r}\varphi _{,r}r^{2}dr
$, where $r_{\bullet }$ stands for the radius of NS $%
r_{s}\left( =\frac{M}{2\gamma }\right) $[ see Eq.(22)] or the throat radius $%
r_{\text{th}}$ of WHs [see Eq.(40), prime dropped], as the case may
be. No Eddington luminosity for SBH since there is no scalar field (in
virtue of no-hair theorem). By using the the scalar field from the metrics
(16,19) and (36,39) respectively, we obtain
\begin{eqnarray}
M^{\text{BNS}}(r)&=&\frac{\pi }{2}\frac{(1-\gamma ^{2})M^{2}}{\gamma ^{2}}
\int_{r_{\bullet }}^{r}\frac{1}{r^{2}}\left( 1-\frac{M}{2\gamma r}\right)^{2(\gamma -2)} \nonumber \\
&&\times\left( 1+\frac{M}{2\gamma r}\right) ^{-2(\gamma +2)}dr,
\end{eqnarray}

\begin{eqnarray}
M^{\text{EBWH}}(r)&=&\frac{8\pi \delta ^{2}M^{2}}{\gamma ^{2}}
\int_{r_{\bullet}}^{r}\frac{1}{r^{2}}\left( 1+\frac{M^{2}}{4\gamma ^{2}r^{2}}\right)^{-4} \nonumber \\
&&\times\exp \left[ 2\pi \gamma -4\gamma \tan ^{-1}\left( \frac{2\gamma r}{M}\right) \right] dr.
\end{eqnarray}

The corresponding coordinate-dependent Eddington luminosity can be obtained
as
\begin{eqnarray}
L_{\text{Edd}}^{\text{BNS}}(r) &=&\frac{1.37\times 10^{33}\times (1-\gamma^{2})M^{2}}{\gamma^{2}} \int_{r_{\bullet }}^{r}\frac{1}{r^{2}} \nonumber \\
&&\times \left( 1-\frac{M}{2\gamma r}\right) ^{2(\gamma -2)}\left( 1+\frac{M}{2\gamma r}\right)^{-2(\gamma +2)}dr  \nonumber \\
&=&1.37\times 10^{33} M^{2}\times l_{\text{Edd}}^{\text{BNS}}(r),
\end{eqnarray}

\begin{eqnarray}
L_{\text{Edd}}^{\text{EBWH}}(r) &=&\frac{1.09\times 10^{34}\times (1+\gamma^{2})M^{2}}{\gamma^{2}} \nonumber \\
&&\times\int_{r_{\bullet }}^{r}\frac{1}{r^{2}}\left( 1+\frac{M^{2}}{4\gamma ^{2}r^{2}}\right)^{-4} \nonumber \\
&&\times \exp \left[ 2\pi \gamma -4\gamma \tan ^{-1}\left( \frac{2\gamma r}{M}\right) \right] dr  \nonumber \\
&=&1.37\times 10^{33} M^{2}\times l_{\text{Edd}}^{\text{EBWH}}(r).
\end{eqnarray}

Fig.10 shows the dependence of the dimensionless Eddington luminosity $L_{\text{edd}}(r)/L_{\text{edd}}$ for a BNS with $M=15M_{\odot}$, for different values of $\gamma$. It is seen that with the increase of $\gamma$ Eddington luminosity decreases sharply, down to zero in the case of SBH (since it does not have a scalar field). Behavior in Fig.11 for EBWH is similar to BNS in Fig.10. At small $\gamma$, Eddington luminosity reaches highest values. Comparing the two plots in Figs.10 and 11, we find that values of $\gamma$ close to $1$ ($\gamma = 0.9$ in the case of BNS and $\gamma = 1.5$ in the case of EBWH), the Eddington luminosity of EBWH is always higher than BNS. However, a completely different picture is observed with decreasing $\gamma$ in the case of BNS. It can be seen from the Fig.10 that for local observers, the Eddington luminosity of EBWH is higher than BNS, but for asymptotic observers, objects such as BNS will be much brighter than EBWH.

\begin{figure}[!ht]
  \centerline{\includegraphics[scale=1.8]{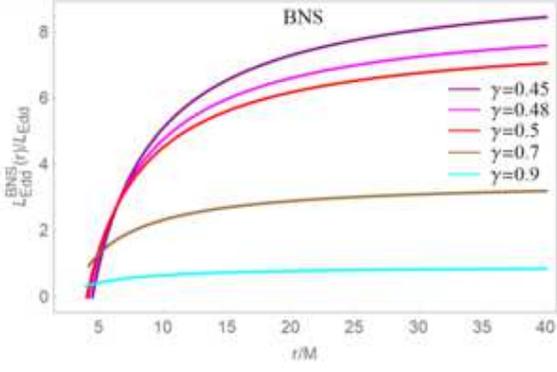}}
  \caption{The Eddington luminosity of BNS for different values of $\protect\gamma$.}
  \label{EddLumBNS}
\end{figure}

\begin{figure}[!ht]
  \centerline{\includegraphics[scale=1.8]{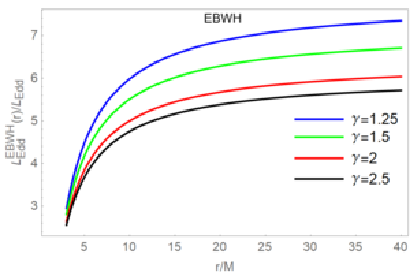}}
  \caption{The Eddington luminosity of EBWH for different values of $\protect\gamma$.}
  \label{EddLumBNS}
\end{figure}

\begin{figure}[!ht]
  \centerline{\includegraphics[scale=1.8]{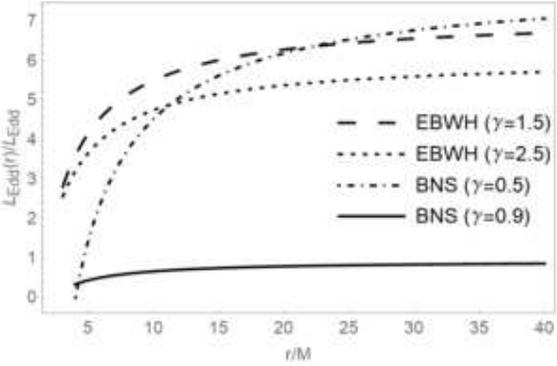}}
  \caption{Comparison of Eddington luminosity of EBWH and BNS for relevant values of $\protect\gamma$ on a logarithmic scale.}
  \label{EddLumBoth}
\end{figure}

The Eddington critical luminosity is the luminosity attained at the critical
radius at which the outward radiation force balances gravity causing
expansion layers of the star. Observation of X-ray bursts\ lead to the
theoretical expectation that the emerging luminosity observed at infinity
should be approximately equal to the Eddington critical luminosity. The
Eddington luminosity at infinity of a NS with a thin surface
is given by \cite{Kovacs:2010}
\begin{equation}
L_{\text{Edd}}^{\infty }=\frac{4\pi m_{p}r_{\bullet }^{2}}{\sigma _{T}}%
\left. \left( \sqrt{-\frac{g_{tt}}{g_{rr}}}\frac{d}{dr}\sqrt{-g_{tt}}\right)
\right\vert _{r=r_{\bullet}}.
\end{equation}

For the BNS, we obtain
\begin{eqnarray}
L_{\text{Edd}}^{\infty \;\text{BNS}} &=& \frac{4\pi m_{p}M}{\sigma _{T}}\left[\frac{16\gamma ^{2}r^{2}}{\left( M^{2}-4\gamma ^{2}r^{2}\right) ^{2}}%
\left( 1-\frac{M}{2\gamma r}\right)^{3\gamma}\right. \nonumber \\
&&\left.\left.\times\left(1+\frac{M}{2\gamma r}\right)^{-3\gamma}\right]\right\vert _{r=r_{s}}.
\end{eqnarray}%
When $r_{s}=\frac{M}{2\gamma }$, we find $L_{\text{Edd}}^{\infty \;(\text{BNS%
})}\rightarrow \infty $. This shows that the Eddington luminosity of BNS can
reach arbitrarily higher values than that of regular astrophysical objects.

For the EBWH we obtain
\begin{eqnarray}
L_{\text{Edd}}^{\infty \;\text{EBWH}}&=&\frac{4\pi m_{p}M}{\sigma _{T}} %
\left[ \frac{16\gamma ^{2}r^{2}}{\left( M^{2}+4\gamma ^{2}r^{2}\right) ^{2}}\right. \nonumber \\
&&\times\exp \left.\left. \left\{ -3\pi \gamma +6\gamma \tan ^{-1}\left( \frac{2\gamma r}{M}\right) \right\} \right] \right\vert _{r=r_{\text{th}}}
\end{eqnarray}%
and after putting $r_{\text{th}}=(\frac{M}{2\gamma })(\gamma +\sqrt{1+\gamma
^{2}})$, we get
\begin{eqnarray}
L_{\text{Edd}}^{\infty \;\text{EBWH}}&=&5.2\times 10^{38}\frac{M^{3}}{M_{\odot}}\left( \gamma +\sqrt{1+\gamma ^{2}}\right)^{2} \nonumber \\
&&\times\exp \left[ -3\pi \gamma+6\gamma \tan ^{-1}\left( \gamma +\sqrt{1+\gamma ^{2}}\right) \right] ,
\end{eqnarray}%
which is in general finite. It can ve verified that $L_{\text{Edd}}^{\infty
\;\text{SBH}}=0$ at $r_{\text{hor}}=\frac{M}{2}$ from Eq.(95) at $\gamma =1$
or from Eq.(96) at $\gamma =i$ [using Eqs.(41) and (42)]. This is consistent
with the fact that SBH has no scalar field exterior.

\section{Conclusions}
\label{sec:concl}
In the foregoing work, we first showed how the NS in the BNS
(or JNW) could be removed by a novel combination of coordinate
transformations, trigonometric identities and complex Wick rotation
converting it to a regular traversible EBWH. The diagram in Fig.1 succinctly
explains how Brans I metric in the JF and BNS, horizonless EBWH, SBH in the
EF are connected among one another. This interesting non-trivial
connectivity does not seem to have been widely noticed as yet, to our
knowledge\footnote{%
In this context, we draw attention to a long standing debate in GR about the
conformally connected JF and EF, viz., which frame is the physical one, and
about the equivalence of the geometrical and physical properties of the
gravitating systems in the two frames. Although we studied accretion
properties exclusively in the EF, we showed earlier \cite{Bhadra:2007} on the basis of
second order light deflection that there are non-equivalent predictions
coming from the two frames, JF and EF. Going a step further,\ predictions of
lensing observables in the two frames also differ, thus they are not
invariant quantities, which is lucidly demonstrated in \cite{Izmailov:2020nn}. A precise
lensing measurement can easily distinguish between the two theories,
supporting one and ruling out the other. The basis for this conclusion is
the fundamental fact that ordinary coordinate transformations do not change
curvature properties, while conformal transformations do change them leading
to corresponding changes in the numerical values of observables. A simple
example is that a flat spaceime can be changed to a non-flat spacetime by
conformal transformation, hence physically observable predictions from them
should be distinct, although light motion in both spacetimes is defined by
the same invariant equation, $d\tau ^{2}=d\widetilde{\tau }^{2}=0$. We thank
the anonymous first referee for raising this important issue.}.

Next we investigated whether, despite being mathematically connected, the
objects BNS, EBWH and SBH, all belonging to EF, can nevertheless exhibit
distinct accretion profiles. We studied their kinematic and emissivity
profiles using the Page-Thorne model applying it to an illustrative
stellar-sized compact accreting object with mass $15M_{\odot }$ and
accretion rate $\dot{M}=10^{18}$ gm.sec$^{-1}$ playing the role of the three
objects in succession. Note that the adopted model is defined explicitly for
$r\geq r_{ms}$ and not for $r\geq r_{s}$ [see Eqs.(76,78,81)], hence we
calculated the profiles from $r_{\text{ms}}$ upwards. Figs.2-4 show
kinematic profiles. Figs.5-7 respectively show accretion flux, temperature
and luminosity that indicate noticeable increase in the BNS profiles as $%
r\rightarrow r_{ms}+$. Figs. 8,9 show differential accretion luminosity
profiles $\frac{d\mathcal{L}_{\infty }}{d\ln {r}}$ on a logarithmic scale.
In particular, Fig.5 (\textit{top panel, right hand}) shows that BNS
accretion flux $F(r)$ is about 6 times higher at the peak than that of the
SBH; Fig.6 (\textit{top panel, right hand}) shows BNS temperature $T(r)$ is
about 2 times higher at the peak than that of the SBH; Fig.7 (\textit{top
panel, right hand}) shows BNS luminosity $\nu L(\nu )$ is about 2 times
higher at the peak than that of the SBH. Of course, nothing can blow up at $%
r=r_{ms}$, where the spacetimes are regular. Still there is noticeable
increase in the emissivity properties of the BNS over the other two regular
objects. However, the accretion efficiency defined by Page-Thorne model \cite{Page:1974}
is $\epsilon =1-\widetilde{E}_{\text{ms}}$, where $\widetilde{E}_{\text{ms}}$
is the specific energy of the accreting particles at the regular limiting
radius $r_{\text{ms}}$. Since $\widetilde{E}\neq 0$ at $r=r_{\text{ms}}^{%
\text{BNS}}$, we obtained according to this model the $\epsilon $ for SBH
the well known value $0.0572$ or just $5.72\%$ efficiency at $r_{\text{ms}}^{%
\text{SBH}}/M=4.949$ deduced as corollaries from BNS ($\gamma =1$) and EBWH (%
$\gamma =i$), and for BNS ($\gamma =0.45$), it is around $8\%$ at $r_{\text{%
ms}}^{\text{BNS}}/M=2.809$ as exhibited in Table 1. These results accord
well with those in \cite{Kovacs:2010}.

It is of interest to note from Eq.(85) that $\widetilde{E}^{\text{BNS}%
}\rightarrow 0$ as $r\rightarrow r_{\text{s}}^{\text{BNS}}=\frac{M}{2\gamma }
$, so that $\epsilon \rightarrow 1$ or $100\%$ efficiency, which agrees with
the conclusion in \cite{Joshi:2014} arrived at by integrating the spectral luminosity,
but this happens only in the singular limit $r\rightarrow r_{\text{s}}^{%
\text{BNS}}$, not covered by the Page-Thorne model. Marginally stable orbits
in BNS cannot not come close the NS. The demand that the
expression for $r_{\text{ms}}^{\text{BNS}}$ in Eq.(82) be real further
reduces the original BNS range $0<\gamma <1$ to the range $\frac{1}{\sqrt{5}}%
<\gamma <1$, which then corresponds to the radial ranges $2.5<$ $r_{\text{ms}%
}^{\text{BNS}}/M<4.949$ and the singular radii $0.5<$ $r_{\text{s}}^{\text{%
BNS}}/M<1.118$. These clearly show that the intervals are \textit{disjoint},
not allowing even a nearly common value $r_{\text{ms}}^{\text{BNS}}/M\sim r_{%
\text{s}}^{\text{BNS}}/M$, so it is \textit{not} possible that minimally
stable orbits could gradually zero in onto the singularity. As a result, an
arbitrarily large increase in the emissivity properties at the singular
radius $r_{\text{s}}^{\text{BNS}}$ cannot be attained within the Page-Thorne
model unlike that in \cite{Joshi:2014}. This happens probably due to the generic
difference between the BNS of a self-consistent theory and the matched NS in
\cite{Joshi:2014} - the former singularity is removable by combinations of non-trivial
transformations, while that in the latter is irremovable. Further, in the
BNS, $r_{\text{s}}^{\text{BNS}}=\frac{M}{2\gamma }$, the center $r=0$ is a
regular point, where scalar curvatures are finite. In the other NS \cite{Joshi:2014},
there is central singularity $r_{\text{s}}^{\text{NS}}=0$, where scalar
curvatures diverge.

Apart from the above differences, the BNS is an EF version of Brans-Dicke
NS threaded by a massless scalar field. This scalar field $%
\phi $, defined in the entire open interval $(r_{s},\infty )$, gives rise to
a very important observable, the \textit{Eddington luminosity}, $L_{\text{Edd%
}}^{\text{BNS}}(r)$. The analyses in Sec.6 lead us to conclude that naked
curvature singularity can exhibit arbitrarily large Eddington luminosity at
infinity $L_{\text{Edd}}^{\infty \;}$ as $r\rightarrow r_{s}$, which are
consistent with the results obtained by Kov\'{a}cs and Harko \cite{Kovacs:2010}.
Interestingly, the gravity fields of massless ($M=0$) BNS and EBWH do
\textit{not} at all accrete matter since $r_{ms}$ is imaginary or infinitely
large respectively, as is shown in Table 1, but both deflect light [see
Eqs.(67) and (70)] since the Keplerian mass\ sensed by the photons is
non-zero ($m\neq 0,\gamma =0$). These cases are important since massless
EBWH has actually been modeled as galactic halo objects in the Milky Way
\cite{Abe:2010,Lukmanova:2016}.

The overall conclusion is that the three objects considered here are in
principle distinguishabe by their accretion properties. In particular,
though not all naked singularities, like BNS here and that in \cite{Joshi:2014}, have the
same efficiency, BNS may still exhibit profiles noticeably higher than those
of non-singular objects in the universe.

\begin{acknowledgements}
The reported study was funded by RFBR according to the research project No. 18-32-00377.
\end{acknowledgements}

\end{document}